\documentclass[journal]{IEEEtran}

\usepackage[pdftex]{graphicx}
\usepackage{wrapfig}

\renewcommand{\arraystretch}{1.5}
\usepackage{booktabs}
\usepackage[table,xcdraw]{xcolor}
\usepackage{adjustbox}
\usepackage{multirow}

\begin{document}
\raggedbottom

\title{ Rethinking IoT Security: A Protocol Based on Blockchain Smart Contracts for Secure and Automated IoT Deployments }

\author{\IEEEauthorblockN{John Wickstr{\"o}m,
Magnus Westerlund, and
G{\"o}ran Pulkkis\\} 
\IEEEauthorblockA{Department of Business and Analytics\\
Arcada Univesity of Applied Sciences\\
Helsinki, Finland\\ Corresponding author: magnus westerlund (at) arcada (dot) fi}}


\maketitle

\begin{abstract}
Proliferation of IoT devices in society demands a renewed focus on securing the use and maintenance of such systems. IoT-based systems will have a great impact on society and therefore such systems must have guaranteed resilience. We introduce cryptographic-based building blocks that strive to ensure that distributed IoT networks remain in a healthy condition throughout their lifecycle. Our presented solution utilizes deterministic and interlinked smart contracts on the Ethereum blockchain to enforce secured management and maintenance for hardened IoT devices. A key issue investigated is the protocol development for securing IoT device deployments and means for communicating securely with devices. By supporting values of openness, automation, and provenance, we can introduce novel means that reduce the threats of surveillance and theft, while also improving operator accountability and trust in IoT technology.
\end{abstract}


\IEEEpeerreviewmaketitle

\section{Introduction}

\IEEEPARstart{D}{EPLOYMENT} of IoT networks and computing on edge nodes have become essential for digitizing and interacting with the physical environment. The expected growth of the IoT sector is rapid \cite{mckinsey}. Over the coming decade there will be billions of newly deployed IoT devices. This development introduces IT architecture changes with challenges to maintain centralized control and proper audit trails to settle independent forensic disputes.

Moving towards distributed systems, without centralized control, makes security-by-design crucial. Current IoT security solutions often have a design based on a traditional master/slave architecture, which also has dominated the general IT security perspective. The master/slave architecture has been instrumental for securing conceptually centralized nodes, i.e. nodes directly supervised by a trusted authority. However, distributed systems need new means to achieve trust. To ensure trusted IoT devices without a direct supervision by a trusted authority, built-in protocol-level control is necessary. Perhaps most important is automating the governance of devices after initialization.

An enabling cryptographic technology for creating a new protocol for autonomous entities is Distributed Ledger Technology (DLT). DLTs usually assume that a community of nodes can together ensure transaction validity. Hence, this displaces the need for a central benevolent authority. Our work focuses on the largest smart contract enabled blockchain, Ethereum. However, we have aimed at creating a generic solution, suitable for any blockchain providing smart contracts. Smart contracts – scripting statements evaluated anonymously by blockchain peers – can in our case specify and enforce rules for access to IoT devices. Smart contracts execute on the Ethereum Virtual Machine (EVM). Blockchain network validators run the EVM and evaluate these contracts for every corresponding transaction, e.g. when a contract function is called. Antonopoulos and Wood \cite{antonopoulos2018mastering} explain smart contract design and use further. 

Utilizing smart contracts as proposed, provides the complete deployment history of an IoT device. The contract transaction records are stored on the blockchain and can provide a blockchain secured auditable log when the device is produced, delivered to an owner, installed, updated, uninstalled, delivered to another owner, reinstalled,  and so on. Thus, smart contracts enable message exchange \cite{conoscenti2016blockchain}. This addresses the current plug-and-play solution design that is often poor, offering attackers numerous attack vectors and vulnerability exploits. Perhaps the worst feature of the plug-and-play design is that owners of infected systems often have no awareness of contamination and forensic discovery is near to impossible \cite{kolias2016learning}. The recently discovered 19 different vulnerabilities, labeled Ripple20, in a popular TCP/IP library amplifies the call for improved IoT security protocols \cite{ripple-zero-day}. The vulnerability discoverer, JSOF research lab, recommend measures that minimize the risk for an attacker gaining complete control over the targeted device remotely. Among others, such calls motivate our reconsideration of the process for achieving secure IoT networks and the proposal of a novel protocol, utilizing smart contracts on a distributed ledger.

The paper has the following structure: section~\ref{section:s2} discusses related smart contract research, section~\ref{section:s3} describes our proposal, section~\ref{section:s4} provides a trust evaluation of the protocol, section~\ref{section:s5} presents the gas consumption associated with smart contract instantiation, and section~\ref{section:s6} offers a discussion and conclusion. 

\section{Other Blockchain Smart Contract-Based Proposals for Secure IoT Deployment}
\label{section:s2}

Restuccia et al. \cite{restuccia2019blockchain} provide a blockchain-based IoT application survey for smart energy, smart environments, robotics, transportation, supply chain, and others. They emphasize that “results show that several IoT applications would benefit from blockchain technologies such as smart contracts”. Several proposed blockchain based solutions for controlling access to IoT devices rely on smart contracts \cite{lin2018secure, novo2018blockchain, xu2018blendcac, zhang2018smart}. 

Anderson et al. \cite{andersen2017wave} present a blockchain-based authorization system called WAVE (Wide Area Verified Exchange) for IoT devices. Embedded agents represent IoT devices, servers, and gateways on the Overlay Layer. Agents and routers interact with the underlying Ethereum blockchain through four smart contracts: WAVE object contract, registry contract, alias contract, and router affinity contract. 

Bahga and Madisetti \cite{novo2018scalable} present a blockchain-based platform called BPIIoT for Industrial IoT (IIoT) where IoT devices are machines utilized in manufacturing environments. BPIIoT enhances the functionality of a recent on-demand manufacturing model called Cloud-Based Manufacturing (CBM) \cite{bahga2016blockchain}, CBM service users issue transactions to smart contracts related to IIoT devices.

Novo \cite{novo2018blockchain, wu2015cloud} proposes an architecture for IoT access management using smart contracts. The solution uses units called management hubs between the IoT network and the smart contracts. This is partially due to resource constraints, to avoid having to synchronize a full blockchain node for every IoT device. Their solution introduces an extra component in the system that constitutes a point of centralization.

Rashid and Siddique \cite{rashid2019smart} compare blockchains with smart contract support and propose the following research directions related to smart contract use in blockchain-based IoT security solutions: resilience against hybrid attacks, optimal IoT system platforms, auditing protocols, management of trust and reliance in social networks, energy-efficient mining, and hybrid consensus protocols.

Biswas et al. \cite{biswas2018scalable} propose a scalable blockchain-based framework for IoT systems, where a local Certification Authority (CA) registers and authenticates local IoT devices. IoT applications use a blockchain Standard Development Kit (SDK) to create smart contracts and transactions. A testbed setup for the proposed framework uses HyperLedger Fabric \cite{hyperledger-fabric}.

Hang and Kim \cite{hang2019design} propose a layered IoT blockchain platform for ensuring integrity of data sensed by IoT devices. Linked IoT devices constitute an IoT Physical Layer. An Application Layer Client executes a smart contract consisting of a function for device owner registration, a device registration function, a sensor reading function, an actuator writing function, and an event notification to update the blockchain ledger.

Huh et al. \cite{huh2017managing} propose the use of Ethereum to manage configuration settings and operation modes of IoT devices. For identification, each IoT generates a unique cryptographic key pair and stores the private key securely. The public key is stored as a transaction record in an Ethereum block and addresses the IoT device. Smart contracts implement IoT device behavior. 

He et al. \cite{he2018privacy} propose a secure blockchain-based management scheme for IoT devices. Transaction records, which can contain smart contracts, are stored in blockchain blocks encrypted with ciphertext-policy attribute-based encryption \cite{bethencourt2007amp} and signed with the Elliptic Curve Digital Signature Algorithm (ECDSA). 

Christidis and Devetsikiotis \cite{christidisblockchains} propose a setup requiring an IoT device manufacturer to provide a master node in a blockchain for IoT firmware updates and to configure all manufactured devices as nodes in the same blockchain network. IoT devices use smart contracts to retrieve hashes of new firmware updates from the blockchain.

Nichol and Brandt \cite{nichol2016co} propose blockchain-based device management for enhanced reliability and security of medical IoT devices. Pre-programmed smart contracts can send automatic notifications about device service requirements to users.

\section{ A Protocol for Distributed IoT Security }
\label{section:s3}

In our review of prior literature, we find many related proposals that consider the use of smart contracts in combination with IoT, e.g. \cite{wu2015cloud} and \cite{he2018privacy}. However, we find none that base their management process (including device handling and event processing) of IoT networks on governance through blockchain technology. We propose a solution for IoT device management that, through smart contracts, enables an operationally autonomous device, i.e. a device self-contained by disallowing any externally initiated incoming network connections. We achieve this firstly (1) by hardening the IoT device to disallow any incoming connections and only allowing certain whitelisted outgoing connections for relevant services, a procedure that significantly reduces attack vector risks. Secondly (2), by utilizing the Ethereum ledger as a mediator layer that provides a task (job) queue, we allow an owner (user) to still maintain the system and command the IoT device/network to perform various actions. Thirdly (3) the device understands the reporting or action backend by independently handling either processing (incl. storing) data or dealing with actions, e.g. through an actuator. Points (2) and (3) further mean that the solution reduces vectors of man-in-the-middle attacks that may occur if the protocol deploys a centralized management node instead of a public blockchain. Point (3) extends to an implementation that enables a participatory platform for task handling, by introducing an associated token based on Ether. 

\begin{figure*}[ht]
    \centering
    \includegraphics[width=0.98\linewidth]{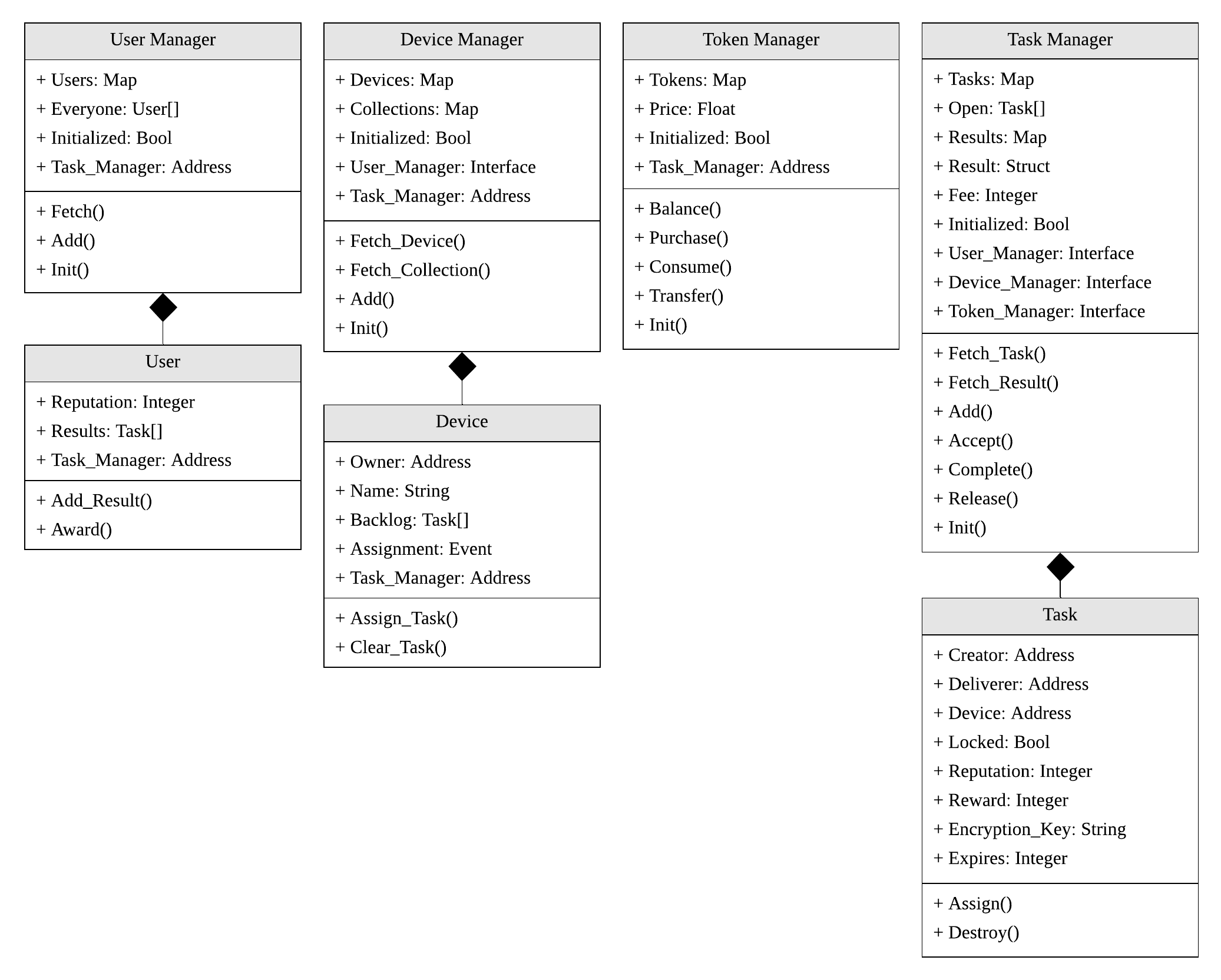}
    \caption{ Smart contract structures and API. Source code and deployment settings at: [https://github.com/wickstjo/eth-task-queue] }
    \label{api-figure}
\end{figure*}

In this paper, we propose a smart contract-based security protocol. There are other parts of the implementation, including a React based Web-UI for interacting with the smart contracts, and a Python application that runs on a Raspberry PI-based IoT device and functions as an interpreter of messages received from the blockchain. However, this paper will primarily focus on the developed smart contract protocol. Figure \ref{api-figure} shows a diagram overview of our smart contracts and indicates important functions and data structures. The development phase has utilized an Ethereum testnet. Any references to Ethereum carry that scope. The contracts are open sourced. Any interested party can publish their own complete contract set, thereby creating an identical but separate backend platform. The solution requires that all users of the system understands secure management of Ethereum private keys. If these keys are lost or compromised, reinstallation is necessary for related IoT devices.

\subsection{ Smart Contract Design }

To achieve trust in a smart contract enabled system, without any governing and trusted intermediaries, all designed transactions must be measurable and deterministic in nature for blockchain peers to be capable to validate them. The implication of this is that an execution of a smart contract should not infer dependencies to externally located data but rather push any data needed for contract evaluation to the ledger. 

In our implementation, we utilize a design with manager contracts that independently verify information and decide whether a process can continue or not. We inject references to other manager contracts via the function init, which effectively binds them together into one coherent system with shared data. Initialization of a manager contract is possible only after deployment of the system. Initialization locks and prevents modifications of static variables like contract references, the token price, or the task fee. By utilizing the “msg.sender” syntax, we limit the right to execute certain functions to manager contracts. This is done to guarantee that necessary checks are performed prior to the execution, which confirm that the other manager contracts consent to the action.

Registered users and devices are stored in key-value maps located in their respective manager contracts. The registration requires a unique identifier for the key, while an an instance of the respective User/Device smart contract will serve as the value. For users, the unique identifier is the Ethereum wallet address. When executing functions on the EVM that write new data onto the ledger, the wallet address of the user cannot be hidden or falsified and can therefore be a trusted identifier.

To create the unique identifier for devices, we chose to hash a concatenation of the serial number embedded in the device chipset and a randomly generated salt. The first execution of the device middleware generates this identifier and stores it locally for subsequent use. The identifier's only requirement is to be unique, so we implemented an easy-to-use naming convention that makes it statistically improbable to produce duplicates. If there is a duplicated device identifier in the contract, then repeated hashing with another salt continues until the generated device identifier is unique. A device owner could also substitute our solution with any random string that is not already in use, when registering their device. For production use, registration may need to further disincentivize “identifier pinching”, i.e. selfishly reserving identifiers that are not needed. 

All system wide functionality becomes available to registered users, except for performing actions on tasks (accepting or completing), which also requires a registered and active device of their ownership. Devices utilize the distributed ledger as a generalized task (job) queue. This means the owner can maintain or command the device to perform various actions without being in direct contact.

\subsection{ Tokenization for Task Handling and Participatory Trust }

The concept of staking, in which each peer puts up a stake for participating in an ecosystem, has gained support in the distributed ledger community. In this regard, tokenization refers to using an arbitrary digital item of value, connected to an underlying crypto coin such as Ethereum’s ether, for providing a stake in the form of a token, in order to be able to participate. Staking should not be confused with the proof-of-stake consensus algorithm. Token staking can ensure that all participants have something to lose if they misbehave and therefore induces some form of participatory trust.

A task consists of two human actor (User) roles, the creator and the respondent, and an IoT device that performs the task on behalf of the respondent. In a base scenario, the creator would assign a task to a known respondent and device (maybe even their own device). We have however implemented an elementary token economy, where registered users can purchase tokens for creating new tasks and provide respondents with token rewards for task completion. Therefore, we also provide a task market platform that enables the creation of public tasks for execution by any respondent’s device. To disincentivize poor conduct, we utilize token staking as a hindrance for unsolicited task creation and task pinching. The respondent must also stake half the token reward offered by the creator of the task, when accepting the task. This prevents malicious respondents blocking other devices from performing the task. This means that our system is also usable for public task allocation and fulfillment, i.e. anyone that meets a set of given criteria can accept such tasks. The task manager contract provides a public registry of available tasks. Even in the base scenario where the creator and respondent are the same, the tokenization serves an important role by creating an immutable audit trail of performed/failed tasks.

For further enforcement of participatory trust, every user contract has a reputation attribute that reflects the user's task participation history. When a task is completed, the creator is rewarded with one reputation point and the respondent with two reputation points. This attribute allows task creators to limit particularly important tasks to respondents who have proven themselves competent, by setting a sufficiently high reputation requirement for the task. In section 6, we outline future research directions based on this initial implementation.

\subsection{ Task Lifecycle }

\begin{figure}[b]
    \centering
    \includegraphics[width=3.45in]{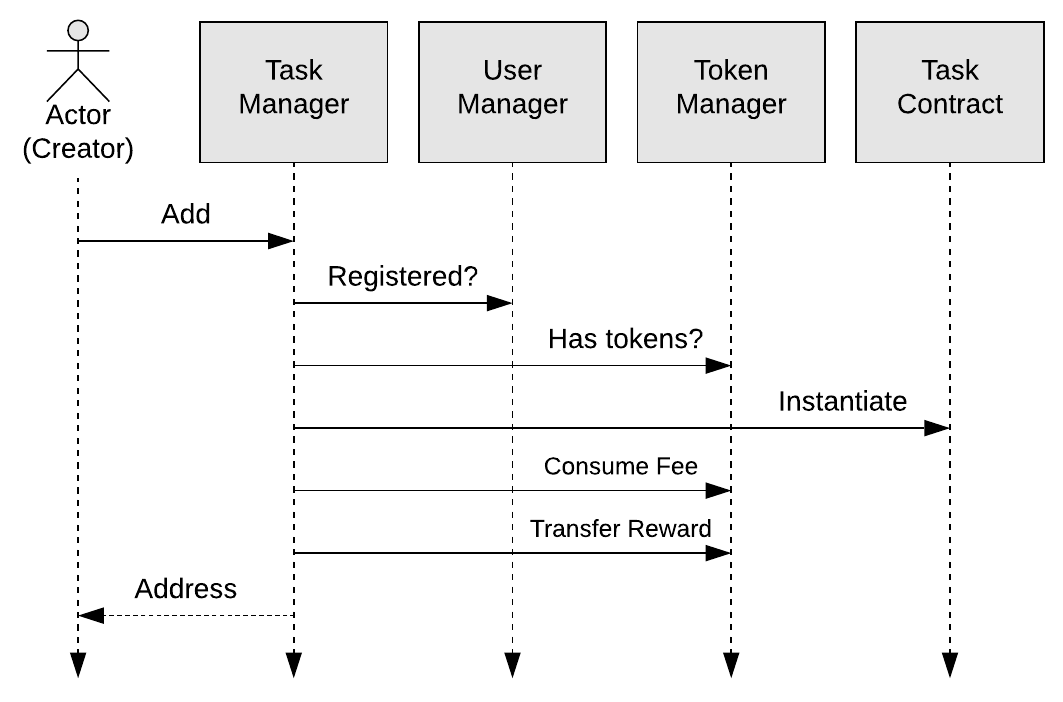}
    \caption{ The Creator instantiates a new Task. }
    \label{create-figure}
\end{figure}

\begin{figure}[t]
    \centering
    \includegraphics[width=3.45in]{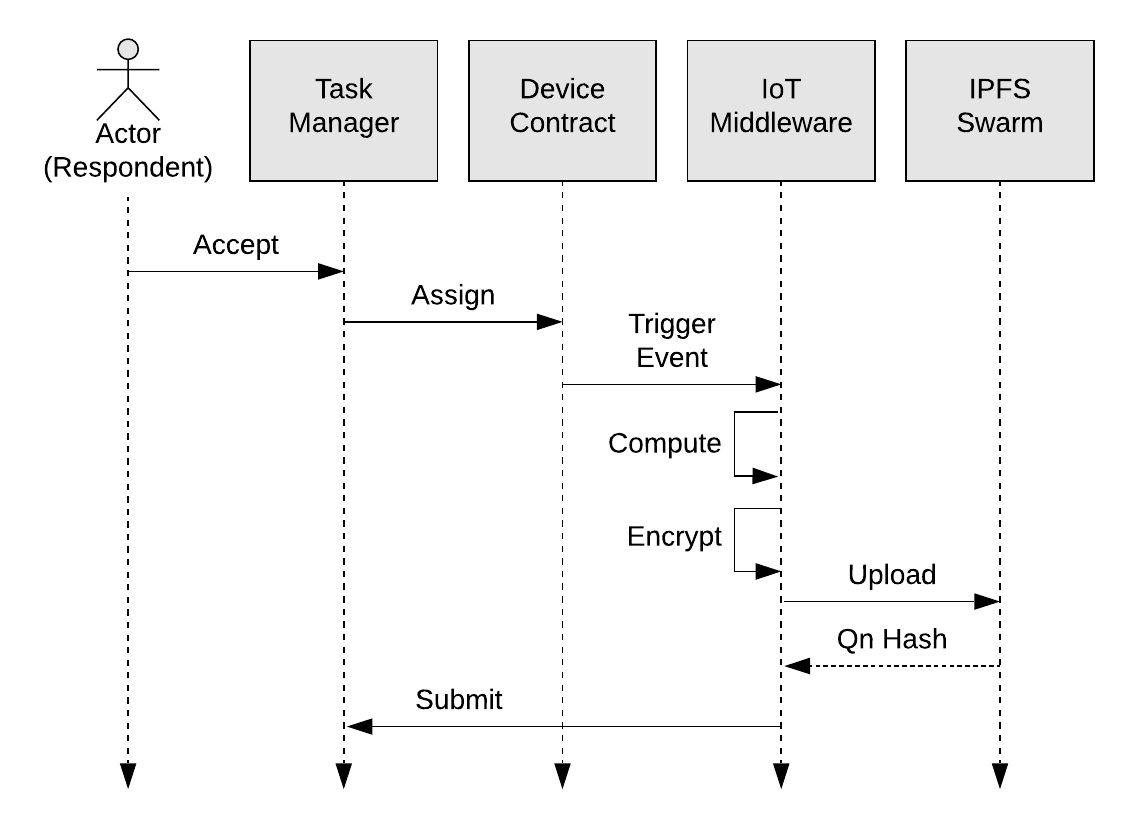}
    \caption{ The Respondent accepts a Task. }
    \label{accept-figure}
\end{figure}

\begin{figure}[ht]
    \centering
    \includegraphics[width=3.45in]{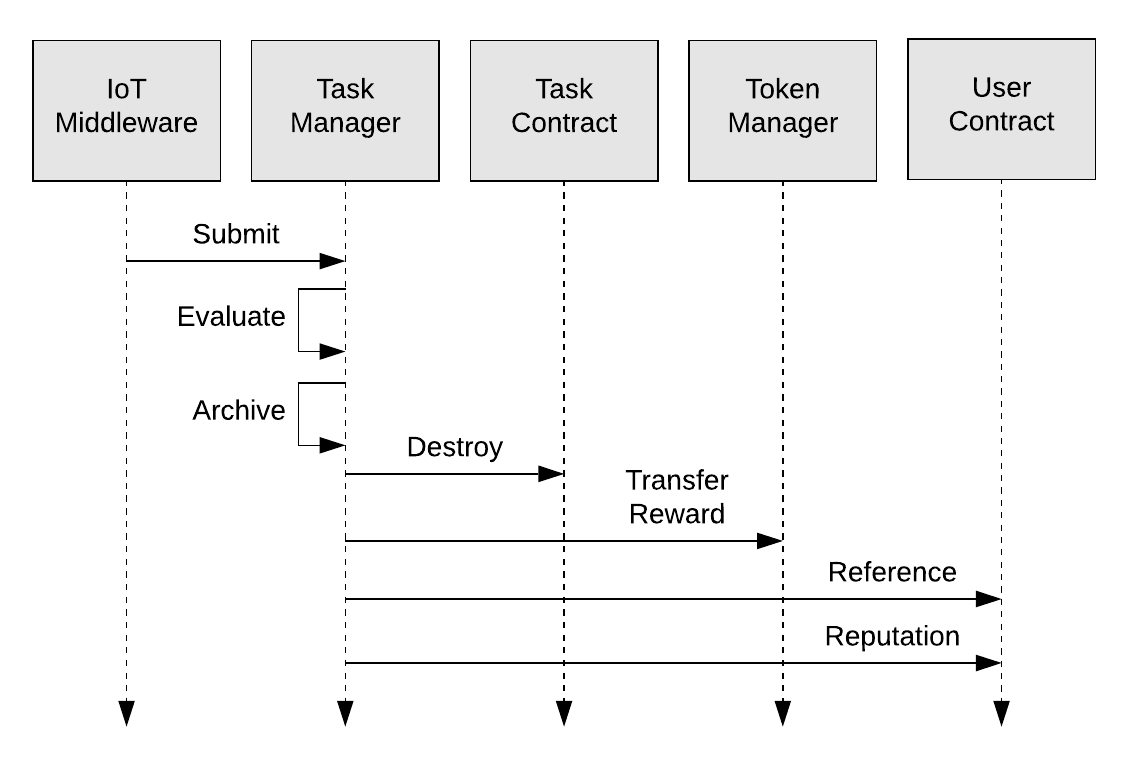}
    \caption{ The task is autonomously completed. }
    \label{complete-figure}
\end{figure}

Tasks are by definition transient and therefore have a different lifecycle compared to user or device contracts. In our proposed solution, tasks are broken down into three separate phases. The creator actor triggers the first phase (see Figure \ref{create-figure}) by instantiating a new task via the task manager. Another actor, the respondent, triggers the second phase by accepting an open task and transferring it to a device backlog – see Figure \ref{accept-figure}. Submission of a valid result to the task manager by a device, or an automatic release of an expired task, triggers the third phase - see Figure \ref{complete-figure}.

Multiple cross-contract verification checks precede the start of a new phase. Any check that returns false will prematurely abort the transaction process and instead return an error message. The token stake that both task actors accept goes to the task itself and remains frozen until the task is resolved as specified by its internal logic.

To accept a task, the respondent must assign a device of their ownership to perform it. The task manager then transfers the task automatically on behalf of the respondent by adding an entry to the device’s backlog in the corresponding smart contract instance. This triggers a custom event, to which a correctly configured device can subscribe. On detection of an event, the device will then autonomously execute the task it has been assigned.

File storage remains an unresolved frontier in the distributed landscape with multiple promising, yet vastly different resolutions. We chose to implement a reporting backend using the InterPlanetary File System (IPFS) \cite{ipfs}, which is a distributed storage network that uses content addressing to fetch files and directories, rather than their location in a file system. IPFS borrows heavily from established torrent technology and similarly allows its users to participate in P2P-networks (swarms) to gain access to content other participants’ hosts. In our proposed solution, all active devices connect to the same IPFS swarm reporting backend. A device encrypts all files produced before sending them to the reporting backend and making them available to the network. To complete the task, the device then delivers a task result object, including the hash of an encrypted file, to the task manager when submitting the result.

For asymmetric message encryption, two public/private key pairs are required. The creator actor generates the first key pair before creating a task and provides the public key as a parameter for the task instantiation. After processing the task, the device generates the second key pair, encrypts the result using the public key provided by the creator and signs the result using the device private key. The task result object contains, in addition to the IPFS address for the encrypted file, a device generated public key, which the task creator can use in conjunction with the original private key to verify the result. This excludes the exchange of private keys and a compromised key only affects a single task.

Since task contracts are destroyed after reaching their third phase, the corresponding result must be stored in another location. We chose to save the data of completed tasks inside the task manager from where it can be fetched on-demand. To notify the creator of a completed task, an entry is pushed to the creator's user contract. On top of functioning as a permanent reference to the task, the entry also triggers an event that a frontend application can listen to and create an asynchronous visual notification for.

Task completion occurs as soon as a permitted device has submitted a validated result or on reaching a specified block number. Upon task destruction, the accumulated token reward is unfrozen and distributed according to a specified ruleset. The smart contract definition of duration to complete a task can be a time interval or block length criteria. However, to track the passage of time with precision on the ledger is a problematic issue. Therefore, we chose the recommended solution to measure time by block distance, i.e. current block + n blocks. Ethereum’s expected block interval is between 10 and 19 seconds, so a task with a block limit of n can be estimated to expire in 14.5 * n seconds (on average) after its inception. We implemented this feature for providing on-chain decision clarity and to give the task creator a way to refund the tokens from an unredeemed task, or if a respondent fails to submit a valid result.

\section{Evaluation of Protocol Trust}
\label{section:s4}

A trust model usually evaluates IoT designs and architectures. Yan et al. \cite{yan42amp} and Sicari et al. \cite{sicari2015security} survey trust models for IoT systems. Asiri \cite{asiri2018blockchain} and Pietro et al. \cite{di2018blockchain} propose trust models for blockchain-secured IoT systems. Kouzinopoulos et al. \cite{kouzinopoulos2018implementing} present a user trust model for a blockchain-secured network of smart homes with various connected IoT systems. The model is a distributed application interacting with smart contracts on a private blockchain. 

Trust evaluation of our proposed IoT security protocol uses IoT trust management taxonomies presented in \cite{yan42amp}. The IoT and the distributed ledger community have a somewhat different definition of the term trust. Ledger-based trust is based on a cryptographic proof and can therefore be considered as a binary true/false value.  Our trust model deploys trust in the sense of excluded or highly improbable incorrectness, as is normal in IoT systems design. In some cases, trust is therefore a form of confidence. Table 1 presents an overview of our protocol’s trust model in relation to current understanding of each sub-topic.

\begingroup
\renewcommand{\arraystretch}{2}

\begin{table*}[t]
    \centering
    \caption{ Trust model overview }
    \begin{adjustbox}{width=6.8in}
    \begin{tabular}{|l|l|l|}
        \hline
        
        \rowcolor[HTML]{CECECE} 
        CONCEPT & DESCRIPTION & IMPLEMENTATION
        \\ \hline
        
        Data perception trust
        
            & Trusted computing, intrusion prevention
            & IoT devices hardened to disallow incoming connections
            \\ \cline{2-3}
            
            & Device fault detection
            & Outlier detection as unexpected deviations from maintained averages
            \\ \hline
    
        Identity trust and privacy preservation
        
            & Identification of users and devices
            & Ethereum based authentication                                             
            \\ \cline{2-3}
         
            & Zero trust model
            & Authentications and transactions logged on the Ethereum ledger            
            \\ \cline{2-3}
            
            & Privacy
            & No personal data stored on the public Ethereum ledger                     
            \\ \cline{2-3}
             
            & Privacy preservation
            & End-to-end encryption of transmitted data                                 
            \\ \hline
        
        Transmission and communication trust
        
            & Not eavesdropped, not modified
            & Lightweight symmetric encryption of transmitted data                      
            \\ \cline{2-3}
            
            & Only authorized communication
            & Asymmetrically encrypted  encryption key exchange, digital signatures     
            \\ \cline{2-3}
            
            & Tokenized protection
            & Token staking in transfer of tasks and task execution results                                            
            \\ \cline{2-3}
            
            & Unblocked communication
            & Detection/Prevention of DoS and DDoS                                      
            \\ \hline
        
        Secure multi-party computation
        
            & Secure computation among untrusted parties
            & Output correctness verification, each party knows only own output         
            \\ \cline{2-3}
            
            & Tokenized protection
            & Token staking of task creators, task respondents, and task executing devices         
            \\ \hline
        
        User trust
        
            & Trustworthy IoT devices and services
            & Maintained reputation of IoT devices and services                                 
            \\ \cline{2-3}
             
            & Open source software based services
            & Publicly available service source code and smart contract code                    
            \\ \cline{2-3}
            
            & Sufficient reputation of IoT devices and services
            & Registration of the behaviour of IoT devices and services                          
            \\ \hline
        
        IoT application trust
            
            & Satisfaction of trust mangement objectives
            & Open source smart contracts, privacy preservation, ledger syncing delegation
            \\ \hline
    
    \end{tabular}
    \end{adjustbox}
\end{table*}
\endgroup

\subsection{Data Perception Trust}

The ability to trust data recorded from a distributed IoT device can include a focus on sensors, local caching, edge analytics, and intrusion attacks that aim to alter data or the data collection process. Our protocol for distributed IoT security focuses on hardening IoT devices to disallow new incoming connections. This hardening prevents intrusion attacks from a distance on IoT devices \cite{ukil2011embedded}. 

Outliers represent faults in data gathered from IoT devices \cite{javed2012automated}. An outlier is an unexpected deviation from a maintained average. Our protocol checks that the values of data items gathered from IoT devices are within pre-set limits. Physical access to a device may still enable sensor tampering, but in our system of IoT devices, encryption on an operating system level prevents direct data manipulation attacks in most cases.

\subsection{Identity Trust and Privacy Preservation}

Ledger transactions to the UserManager and DeviceManager contracts register Users and their IoT devices. Registration instantiates and stores User and Device contracts on the Ethereum blockchain. Ethereum based authentication therefore creates trust in user and device identities. Identity trust is a Zero Trust model \cite{kindervag2016no} since network traffic is untrusted, access control enforced by smart contracts verifies identities, and inspection and logging of network traffic takes place on the Ethereum blockchain. To ensure privacy preservation, only content hashes and public encryption keys are stored on the Ethereum ledger and transmitted data is end-to-end encrypted between the device and the reporting backend.

\subsection{Transmission and Communication Trust}

Trusted data transmission and communication is authorized, unmodified, unblocked, and cannot be listened in on. Each transaction sender and transaction receiver generate their own private/public key pair using public key cryptography. A trusted transaction initiated by a sender is both encrypted and authorized. Authorization is a signature created with the sender’s private key. The public key of the transaction receiver encrypts the symmetric key required for decryption of the transaction \cite{isa2012lightweight}. Moreover, trusted management of the public key of the sender and trusted routing in the transmission path to the receiver are additional transaction trust requirements \cite{liu2010trust}. 

Token staking by task creators and task respondents in our protocol induces task transmission trust and trusted transfer of task execution results. Transmission and communication trust also require detection and prevention of DoS and DDoS attacks \cite{heer2011security}, something we achieve by hardening IoT devices to disallow incoming connections and using the ledger as an intermediary between an IoT device and a potential attacker.

\subsection{Secure Multi-Party Computation}

Secure multi-party computations refer to parties, who do not have trust in each other and participate in computations with their own secret inputs. A multi-party computation is by definition trusted when the output is correct, and each party can receive and know only its own output \cite{yan42amp}. 

We introduce a definition for multi-party computation of management trust for two actors and a device. Token staking of task creators, task respondents, and task executing devices in our protocol is a multi-party computation, which induces task management trust. In section 6 we further outline future research for how complex task result validation can be enforced.

\subsection{User Trust}

User trust in IoT devices and in services using IoT devices depends on the reputation of these devices and services. Reputation registers past behavior of devices and services. Ruohomaa et al. \cite{ruohomaa2007reputation} surveys reputation management and Braga et al. \cite{braga2018survey} presents an overview of proposed reputation models. Liu et al. \cite{liu2014trust} propose an IoT node behavior-based trust system combining direct trust, recommended trust, and history statistical trust. In our proposal we have primarily chosen direct and historical reputation for user trust. Trust in services using IoT devices also stems from open source code implementation of such services, since any user can then check the correctness of the implementation of such services.

\subsection{IoT Application Trust}

Trust in IoT applications depends on how well these applications satisfy trust management objectives, such as privacy preservation, data fusion and mining trust, data transmission and communication trust, and identity trust \cite{yan42amp}.

IoT application trust of our protocol stems from open source smart contracts, an operating system encrypted file system, the absence of hard-coded private user keys, private device keys stored in operating system encrypted files, and ledger syncing delegation to Cloudflare's Ethereum Gateway \cite{cloudflare-gateway}. Through this device resource demands are lower and, compared to a private gateway, have increased reliability by using the Cloudflare edge network.

\section{Contract Deployment, Verification, and Gas Consumption}
\label{section:s5}

The smart contracts are written in the Solidity language and converted to bytecode using the 0.6.8 solc compiler \cite{solidity}. Initial development was performed on a local blockchain for debugging purposes, and later using Ethereum’s Ropsten testnet \cite{ropsten}. The Ethereum community maintains a smart contract specific software weakness classification (SWC) scheme and registry \cite{swc-registry}. Static analysis tools utilize the registry of test cases to locate potential security vulnerabilities in the contracts. To test our smart contracts the MythX analysis tool was used \cite{mythx}, which could not find any of the critical vulnerabilities listed in the SWC Registry \cite{swc-registry}. 

In Table \ref{gas-table}, we present the static gas values associated with contract instantiation, and the estimated dollar value they represent. Per our design, manager contracts are only instantiated once, while the child contracts are instantiated per usage. Note that the price of gas fluctuates frequently because it operates via supply and demand, and that calculation of the dollar representation in Table \ref{gas-table} has used an average estimated gas price of approximately 0.4787 USD per 100k gas. 

The cost of adding a new user or device is quite negligible. The current price of a task suggests that the immediate use cases will be longer term subscription models or high value tasks. Other distributed ledgers with a lower transaction cost or corporate chains, may be better suited for low value tasks, such as one-off measurements. As the gas cost is directly dependent on the contract complexity, further engineering may be able to optimize the cost structure.

\section{Discussion and Conclusions}
\label{section:s6}

In this paper, we devise a novel protocol for improving security of distributed IoT devices. We explain how IoT devices can integrate smart contract enforced security by utilizing our protocol. We achieve an improved trust model for a reasonable transaction price by utilizing the public Ethereum blockchain. The distributed ledger approach provides the owner with an openness, security, and specific control of each device in the network that would otherwise be difficult to implement. The novelty of our research is that we have implemented a distributed protocol that, in addition to handling authentication and authorization for actors and devices, also establishes a platform for task handling. Thereby any task creator can receive input not only from their own devices, but also from any respondent’s devices. Extending the protocol’s functionality is still possible, but the protocol provides a clear indication that protocols utilizing distributed ledger technology can improve the security of IoT management and control.

The solution makes no assumptions on the trustworthiness of manufacturers or designers of software running on devices. Rather, we assume that trust is hard to establish in a distributed environment. While hardening the devices to disallow all incoming data connections to them, our solution can still offer the interaction needed for proper utilization.

\begin{table}[t]
    \centering
    \caption{ Smart contract gas values and estimated dollar costs. }
    \begin{adjustbox}{width=2.9in}
        \begin{tabular}{| c | c | c |}
        \hline
        \rowcolor[HTML]{CECECE} 
            Smart Contract & Gas Value & Est. Dollar Cost          \\\hline
            User Manager   & 530 579         & \$2.535             \\\hline
            Device Manager & 1 097 206       & \$5.242             \\\hline
            Task Manager   & 3 052 709       & \$14.585            \\\hline
            Token Manager  & 413 560         & \$1.975             \\\hline
            User           & 273 931         & \$1.308             \\\hline
            Device         & 446 652         & \$2.134             \\\hline
            Task           & 554 883         & \$2.651             \\\hline
        \end{tabular}
    \end{adjustbox}
    \label{gas-table}
\end{table}

The focus of future research will be on expanding the system and on providing a more feature complete management functionality that better caters different contexts. Our work has so far focused on a security protocol that includes task handling. We aim to explore both microservice and edge cloud use cases, including improved tokenization and task handling. 

A reason to design smart contracts is to verify an objective truth, but smart contracts are inherently bad at inferring a subjective truth such as deciding on the sufficiency of a task result. Tasks should be verifiable, with a binary true/false response as a confirmation of its completion. For more advanced tasks so called smart contract oracles can be used in the verification process. Oracles are free standing services that can execute special tasks or fetch information that is not inherently available to the blockchain validator. For example, a task instance can via a function call to the smart contracts under which the oracle operates, ask for more information. Although oracles may introduce certain logic ambiguity and a centralized point of failure, we will further investigate their use in improving task validation and as mediators for interoperability services, something that is important for improving participatory trust, usability, and reputation.

\section*{Acknowledgment}

The authors would like to thank Dr Keir Finlow-Bates for his constructive comments.

\ifCLASSOPTIONcaptionsoff
  \newpage
\fi



%
\bibliographystyle{IEEEtran}
\bibliography{document.bib}


%








\end{document}